\documentclass[superscriptaddress, aps, preprintnumbers, twocolumn, amsmath, amssymb, sort&compress, nofootinbib, 10pt, paper, floatfix]{revtex4-2}

\usepackage{amsfonts,amsmath,amssymb,ascmac,bm,tensor, microtype}
\usepackage{fnpct} 
\usepackage{comment}
\usepackage{ifpdf}
\usepackage{slashed}
\usepackage{color}
\usepackage[mathscr]{eucal}
\usepackage[utf8]{inputenc}
\usepackage{cancel}
\usepackage{soul}
\usepackage{braket}
\ifpdf
  \usepackage{graphicx}     
  \usepackage[bookmarksopen,colorlinks=true,linkcolor=bblue,citecolor=bblue,urlcolor=ppink]{hyperref}
\else     
\fi

\definecolor{red}{rgb}{1,0,0}
\definecolor{darkred}{rgb}{0.6,0,0}
\definecolor{darkgreen}{rgb}{0.992447,0.623778,0.034597}
\definecolor{ppink}{rgb}{1,0.4,0.4}
\definecolor{bblue}{rgb}{0.284602,0.317763,0.963947}
\definecolor{purple}{rgb}{0.5 ,0, 0.7}

\definecolor{dgreen}{rgb}{0 ,0.5, 0.5}

\makeatletter
\newcommand\footnoteref[1]{\protected@xdef\@thefnmark{\ref{#1}}\@footnotemark}
\makeatother

\allowdisplaybreaks[1]

\begin{document}

\title{Probing Long-Range Forces Between Neutrinos with Cosmic Structures}
\date{\today}

\author{David E. Kaplan}%
\author{Xuheng Luo}
\author{Surjeet Rajendran}%
\affiliation{The William H.~Miller III Department of Physics and Astronomy, The Johns Hopkins University, Baltimore, MD  21218, USA}

\begin{abstract}
\noindent We study the consequences of new long-range forces between neutrinos on cosmic scales. If these forces are a few orders of magnitude stronger than gravity, they can induce perturbation instability in the non-relativistic cosmic neutrino background in the late time universe. As a result, the cosmic neutrino background may form nonlinear bound states instead of free-streaming.  The implications of the formation of nonlinear neutrino bound states include enhancing matter perturbations and triggering star formation. Based on existing measurements of the matter power spectrum and reionization history, we place new constraints on long-range forces between neutrinos with ranges lying in $1 \text{ kpc}\lesssim m_\phi^{-1} \lesssim 10 \text{ Mpc}$.
\end{abstract}
\maketitle

\section{Introduction}
The non-zero masses of neutrinos imply the existence of particles beyond the Standard Model, such as right-handed neutrinos. Neutrinos could also serve as a portal to dark sectors, potentially connecting to dark matter, dark energy, and matter asymmetry \cite{Mohapatra:2005wg, Abazajian:2012ys, Drewes:2013gca, Batell:2017cmf, Bertoni:2014mva, Graham:2017hfr, Blennow:2019fhy, Graham:2019bfu, Berghaus:2020ekh,Holst:2023hff}. This motivates extensive searches for neutrino physics beyond the Standard Model in astrophysics, cosmology, and laboratory experiments \cite{Proceedings:2019qno,Babu:2019iml,Dev:2022bae,Luo:2020sho,Brinckmann:2020bcn,Luo:2020fdt,Das:2024thc,Chauhan:2024nfa,Chauhan:2023sci,Wang:2023csv,Wu:2023twu,Li:2023puz,Chauhan:2024qew,Chauhan:2024deu,Craig:2024tky,Green:2024xbb,Wise:2018rnb,Smirnov:2019cae,Ciscar-Monsalvatje:2024tvm,Herrera:2024upj,Loverde:2022wih,Fiorillo:2022cdq,Fiorillo:2023cas,Fiorillo:2023ytr}. Typically, a new light bosonic particle coupled to Standard Model particles can induce long-range forces between atoms, which are strongly constrained by fifth force searches and equivalence principle tests \cite{Will:2014kxa}. However, if a new light bosonic particle is coupled exclusively to neutrinos, the induced fifth forces between atoms are highly suppressed \cite{Xu:2020qek, Chauhan:2020mgv}. Furthermore, the feeble interaction between neutrinos and Standard Model particles not only makes neutrinos difficult to detect, but also makes them very non-localized. Consequently, laboratory searches for long-range forces between neutrinos are very limited.

From a different perspective, neutrinos are a fundamental component of the universe. The standard cosmological model predicts an enormous amount of relic neutrinos in the universe today, known as the cosmic neutrino background (C$\nu$B). In the early universe, these were free-streaming relativistic particles that occupied a significant fraction of the total energy density of the universe. 
They leave imprints on many cosmological observables through gravitational interactions \cite{Bashinsky:2003tk,Baumann:2015rya,Baumann:2017lmt,Green:2020fjb,Follin:2015hya,Hotinli:2023scz,Worku:2024kwv,Bertolez-Martinez:2024wez,Loverde:2024nfi}, which are measured in the cosmic microwave background (CMB) \cite{Planck:2018vyg}, the relic abundance from big bang nucleosynthesis (BBN) \cite{Cyburt:2015mya}, and the large-scale structure of the universe (LSS) \cite{Baumann:2019keh}. In the late time universe, the C$\nu$B should start to become non-relativistic at redshifts around $ 120 \left(\frac{m_\nu}{60\text{ meV}}\right)$ and then its energy density redshifts like matter. This non-relativistic but high-velocity matter leaves imprints on the growth of structure in the universe \cite{Lesgourgues:2006nd,Wong:2011ip,Lesgourgues:2012uu,Hu:1997mj,Kaplinghat:2003bh,DESI:2024mwx}. Galaxy surveys aim to measure this effect with high signal-to-noise ratios to determine the sum of neutrino masses in cosmology \cite{Dvorkin:2019jgs}. These measurements of the C$\nu$B are sensitive to the properties of neutrinos on cosmic scales, placing strong constraints on the physics of the neutrino sector \cite{Esteban_2021,Esteban:2022rjk,Beacom:2004yd,Chacko:2019nej,Chacko:2020hmh,Berryman:2022hds,FrancoAbellan:2021hdb,Green:2021gdc,Bansal:2024afn,Lorenz:2018fzb,Lorenz:2021alz,Kamionkowski:2024axz,He:2023oke,Kreisch:2019yzn,Kreisch:2022zxp,Cyr-Racine:2013jua,Archidiacono:2013dua,Lancaster:2017ksf,Camarena:2023cku,Camarena:2024zck}.

The absence of strong laboratory constraints motivates us to consider the consequences of new long-range forces between neutrinos on cosmic scales. Indeed, the cosmological influence of long-range forces between neutrinos has been considered in the past \cite{Fardon:2003eh,Kaplan:2004dq,Brookfield:2005bz,Franca:2009xp,Gogoi:2020qif,Wintergerst:2009fh,Pettorino:2010bv,Casas:2016duf,Esteban_2021,Esteban:2022rjk,Gogoi:2020qif,Frieman:1991tu}. In these models, a non-trivial impact on cosmology usually requires long-range forces between neutrinos that are much stronger than gravity. However, at the same time, as pointed out in \cite{Afshordi:2005ym}, and subsequently in \cite{Bjaelde:2007ki}, these models generically have perturbation instability, leading to the formation of nonlinear structures. The possibility that neutrinos exist in the form of bound states was studied earlier in Ref.~\cite{Stephenson:1996qj}, then later in Refs.~\cite{Smirnov:2022sfo,Brouzakis:2007aq}.

Here, we study the impact of long-range forces between neutrinos on cosmic structures mediated by an ultralight scalar field $\phi$. In the presence of this interaction, the C$\nu$B remains consistent with free-streaming relativistic relics until the late time universe, and thus is not constrained by early universe observations like the CMB or BBN. However, as the C$\nu$B becomes non-relativistic, the long-range forces between neutrinos induce the Jeans instability in the density field of the C$\nu$B, leading to rapid, non-gravitational perturbation growth. This is evident in both the numerical calculation and the analytical approximation of the modified Boltzmann equation for the C$\nu$B. Consequently, as the C$\nu$B density perturbations become nonlinear, instead of continuing as free-streaming particles, the C$\nu$B forms nonlinear bound states.

In the standard cosmological model, matter density perturbations remain smooth until the late time universe. The early formation of massive, nonlinear structures can enhance the growth of structure. This has been studied in the context of primordial black holes (PBHs) \cite{Afshordi:2003zb,Murgia:2019duy,Inman:2019wvr,Delos:2024poq,Carr:2018rid,Liu:2022bvr,Liu:2022okz,Liu:2023pvq,Zhang:2023hyn,Zhang:2024ytf} and massive compact halo objects (MACHOs) \cite{Bai:2020jfm,Croon:2024rmw,Croon:2024jhd,Irsic:2019iff,Chang:2024fol,Amin:2022nlh}. At large scales, the gravitational influence of these neutrino bound states can be described in a similar way. In doing so, we find a substantial amplification of the matter power spectrum, which can be probed by cosmological surveys. Moreover, these neutrino bound states can trap baryons, and the trapped baryonic gas can cool efficiently, enabling star formation. The early formation of stars within these bound states can significantly alter the reionization history, which is strongly constrained by CMB observations.

The rest of this paper is structured as follows. In Sec.~\ref{sec:model}, we specify the interaction and analyze the cosmological evolution of the C$\nu$B in our model. In Sec.~\ref{sec:cos}, we describe the impact on cosmic structures and compare them with observations. Finally, we discuss directions for future exploration and conclude in Sec.~\ref{sec:con}.

\section{\label{sec:model} Model and Analysis}
We consider the following Lagrangian describing the interaction between a Dirac neutrino $\nu$ with mass $m_\nu$ and an ultralight scalar field $\phi$ with mass $m_\phi$:
\begin{equation}\label{lagrangian}
\mathcal{L} \supset \frac{1}{2}D_\mu \phi D^\mu \phi - \frac{1}{2}m_\phi^2\phi^2 + i\bar{\nu} \slashed{D}\nu - m_\nu\bar{\nu}\nu - g\phi\bar{\nu}\nu.
\end{equation}
Here, $D_\mu$ represents the covariant derivative in an expanding universe. For the purposes of this paper, we focus on $\phi$ coupling to one neutrino, and we briefly discuss extensions to multiple neutrinos in Sec.~\ref{sec:con}.

Throughout this work, we require that the coupling strength is sufficiently small ($g \ll 10^{-7}$) to ensure that collisions and annihilations between neutrinos are negligible on cosmological scales (see also Ref.~\cite{Berryman:2022hds} for a recent review of the cases where these interactions are important).

By default, we set the neutrino mass to $m_\nu = 60 \text{ meV}$ in our analysis. In the late time universe, the fraction of the total energy of matter in the C$\nu$B is
\begin{equation}
f_\nu \equiv \frac{\Omega_\nu}{\Omega_m} \approx 0.45 \% \left(\frac{ m_\nu}{60 \text{ meV}} \right).
\end{equation}

\subsection{Background evolution}

In the presence of the C$\nu$B, $\phi$ is coherently sourced from the interaction in Eq.~\eqref{lagrangian}. Simultaneously, the neutrinos acquire an effective mass $m_\nu + g\phi$. As a result, the C$\nu$B and the background $\phi$ are dynamically coupled. The non-trivial cosmological evolution of this coupled system has led to many studies \cite{Fardon:2003eh,Brookfield:2005bz,Franca:2009xp,Gogoi:2020qif,Esteban_2021,Esteban:2022rjk} (see also Refs.~\cite{Savastano:2019zpr,Amendola:2017xhl,Domenech:2023afs,Flores:2020drq,Flores:2023zpf,Archidiacono:2022iuu,Kesden:2006zb,Kesden:2006vz,Keselman:2009nx,Bottaro:2023wkd,Bottaro:2024pcb,Bogorad:2023wzn,Bogorad:2024hfj} for studies on an ultralight scalar field coupled with dark matter). As for the Lagrangian in Eq.~\eqref{lagrangian}, its background evolution has been studied in Ref.~\cite{Esteban_2021} for large $m_\phi$. We find that in the parameter space of interest, the background evolution is still similar to that in Ref.~\cite{Esteban_2021}.

At the homogeneous background level, the equation of motion (EoM) for the scalar field is
\begin{equation}\label{phibg}
       \Ddot{\phi}_0 + 3H\Dot{\phi}_0+ m_\phi^2\phi_0 = -g\braket{\bar{\nu}\nu}_0(\phi_0),
\end{equation}
where the dot denotes derivatives with respect to the proper time, and we use subscript ``$_0$" to denote the homogeneous component of the field. The full expression of $\braket{\bar{\nu}\nu}$ can be found in Eq.~\eqref{source}. We are interested in the parameter space where 
\begin{equation}\label{criterion2}
      g \gtrsim \frac{m_\nu}{m_{\text{pl}}\sqrt{f_\nu}} \approx 4\times 10^{-28},
\end{equation}
here $m_{\text{pl}}$ is the reduced Planck mass. Otherwise, the interaction is too weak for the $\phi_0$ sourced by neutrinos to have a non-negligible impact on the C$\nu$B.

\begin{figure}
\includegraphics[width=\linewidth]{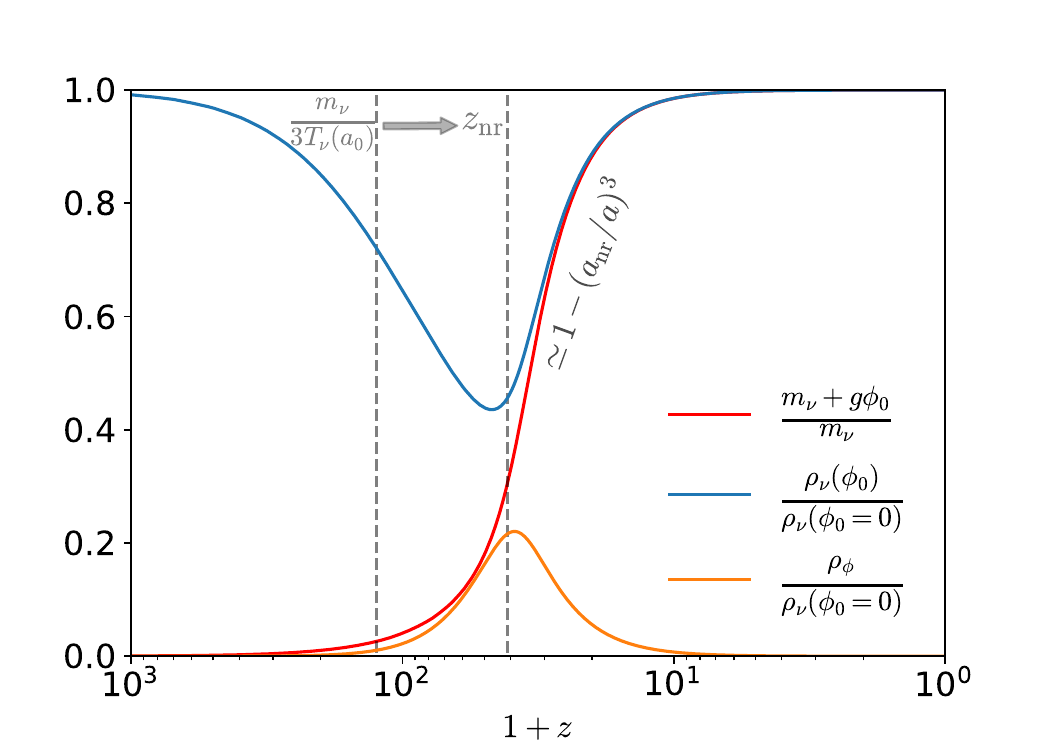}
\caption{\label{fig:epsart} The background evolution of effective neutrino mass, neutrino energy density, and scalar field energy density as functions of redshift. The model parameters are set to $g =10^{-26}$ and $m_\phi = 10^{-29} \text{ eV}$. The vertical dashed lines indicate the redshifts at which the C$\nu$B becomes non-relativistic: the left line corresponds to the $\Lambda$CDM universe, where the transition happens at redshifts $z_{\text{nr}} \approx 120$, and the right line corresponds to our model, where $z_{\text{nr}} \approx 40$ for the chosen model parameters here.}\label{bkg_plot}
\end{figure}

The evolution of $\phi_0$ is illustrated in Fig.~\ref{bkg_plot}. Similar to Ref.~\cite{Esteban_2021}, in the early universe, the effective source term $g\braket{\bar{\nu}\nu}_0$ in Eq.~\eqref{phibg} dominates the evolution of $\phi_0$. A $\phi_0$ is sourced from the background C$\nu$B and is dynamically misaligned to $\phi_0^* \approx -m_\nu/g$. In this period of evolution, as long as the number density of the C$\nu$B is large, the effective mass of neutrinos remains suppressed by $\phi_0$, with $m_\nu +g\phi_0 \approx 0$. As the universe expands, the number density of the C$\nu$B decreases. Eventually, the bare potential term $m_\phi^2\phi_0$ in Eq.~\eqref{phibg} starts to dominate the evolution of $\phi_0$ at redshift $z_{\text{nr}}$, when $gn_{\nu,0}(z_{\text{nr}}) \approx -m_\phi^2\phi_0^*$. After this point, the neutrinos become non-relativistic and $\phi_0$ rapidly decreases at the rate $\propto a^{-3}$, and the effective mass of neutrino returns to its bare mass.

The dynamics of $\phi_0$ result in an evolving neutrino mass, which delays the transition of the C$\nu$B from relativistic to non-relativistic until a lower redshift $z_{\text{nr}}$, where
\begin{equation}\label{eq:nr}
a_{\text{nr}} \approx \max \left[\left(\frac{g^2n_{\nu,0}(a_0)}{m_\phi^2m_\nu}\right)^{1/3},\ \frac{3T_{\nu}(a_0)}{m_\nu}\right],
\end{equation}
here, $a_{\text{nr}} = (1+z_{\text{nr}})^{-1}$, $a_0 \equiv 1$ is the scale factor today, $n_{\nu,0}(a_0) \approx 112\text{ cm}^{-3}$ and $T_\nu(a) \approx 1.95a^{-1}\text{ K}$. In the standard cosmological model, the C$\nu$B becomes non-relativistic at redshifts $(3T_{\nu}(a_0) /m_\nu)^{-1} \approx 120$. In the model parameter space $(g, m_\phi)$, the Yukawa interaction begins to affect $z_{\text{nr}}$ if
\begin{equation}\label{criterion1}
    \left(\frac{g^2n_{\nu,0}(a_0)}{m_\phi^2m_\nu}\right)^{1/3} \gtrsim \frac{3T_{\nu}(a_0)} {m_\nu}\  \Rightarrow \   g \gtrsim 12 \frac{m_\phi}{m_\nu}.
\end{equation}
Otherwise, $\phi_0$ decreases while the C$\nu$B is relativistic, and the C$\nu$B evolution is not significantly affected throughout. We confine our analysis to the regime where both Eqs.~\eqref{criterion2} and \eqref{criterion1} hold. These two criteria also ensure that $m_\phi \gtrsim H(z_{\text{nr}})$, such that the Yukawa interaction range has entered the horizon before the neutrinos become non-relativistic.

In addition to the above leading-order description of the evolution of $\phi_0$, when numerically solving Eq.~\eqref{phibg}, there are also oscillations of $\phi_0$ around the (effective) potential minimum of $\phi$. In the late time universe, these do not significantly change the evolution of neutrino mass for the relevant parameter space, and we therefore ignore them in our analytical model.

The evolution of the effective neutrino mass does not significantly affect the early universe ($z \gtrsim 1100$). This is because, in the early universe, as one can see from Fig.~\ref{fig:epsart}, the energy density of neutrinos is dominated by the relativistic kinetic energy, which is nearly independent of the neutrino mass. Additionally, the energy density of the scalar field is negligible at that time. At the perturbation level, we also find that the evolution of the C$\nu$B perturbations is not strongly affected in the early universe (see Fig.~\ref{fig1}). Therefore, early universe observables such as the CMB and BBN do not directly constrain this model.

Nevertheless, in the late time universe, both the energy density of the C$\nu$B and its transition from relativistic to non-relativistic depend on the neutrino mass. Therefore, similar to the standard $\Lambda$CDM universe, effectively changing the neutrino mass can leave imprints on late time cosmological observables \cite{Lorenz:2018fzb,Lorenz:2021alz,Esteban_2021,Bertolez-Martinez:2024wez}. Moreover, disrupting the free-streaming nature of the C$\nu$B can also impact the formation of structures in a way similar to varying the neutrino mass \cite{Esteban_2021,Green:2021gdc}. However, the influence of these processes is limited by the total energy density of the C$\nu$B in the late time universe, and thus will only induce fractional changes of order $\mathcal{O}\left(f_\nu \right)$ in cosmological observables. Additionally, as studied in Refs.~\cite{Esteban_2021,Esteban:2022rjk}, when fitting the model with observations, these changes can degenerate with the model parameters and other cosmological parameters. As a result, for a small neutrino mass, such as $m_\nu = 60 \text{ meV}$ considered here, it is difficult to probe these effects using the current observational sensitivities, and we will not focus on them for the purpose of this work.

\subsection{Perturbation evolution: the Jeans instability}\label{sec:lpt}
In the presence of new long-range forces stronger than gravity, perturbation instability can occur in non-relativistic cosmic fluids which eventually leads to the formation of nonlinear structure. This has been studied for dark matter \cite{Amendola:2017xhl,Savastano:2019zpr,Domenech:2023afs,Flores:2020drq,Flores:2023zpf} and the C$\nu$B \cite{Esteban_2021,Esteban:2022rjk} (including models of mass-varying neutrinos \cite{Afshordi:2005ym,Bjaelde:2007ki,Wintergerst:2009fh,Pettorino:2010bv,Casas:2016duf}). In our case, the perturbation instability can be intuitively understood through the Jeans criterion, which states that collapse occurs when the potential energy of a cloud exceeds its kinetic energy. For a cloud of non-relativistic C$\nu$B with size $R$ and characteristic velocity $T_\nu/m_\nu$ ($3T_\nu \approx \braket{p_\nu}$), the Yukawa potential energy is larger than the kinetic energy when
\begin{align}\label{eq:prejeans}
    \frac{g^2N_\nu^2}{R}  \gtrsim N_\nu \times \frac{T_\nu^2}{m_\nu}  
    \Rightarrow \  g \gtrsim \sqrt{\frac{T_\nu^2 }{m_\nu n_\nu R^2}},
\end{align}
where we used $N_\nu \sim n_\nu R^3$. The largest $R$ before the above approximation breaks down is the Yukawa interaction range $m_\phi^{-1}$. Substituting $R \sim m_\phi^{-1}$ back into Eq.~\eqref{eq:prejeans} yields the minimum $g$ that triggers Jeans instability in the C$\nu$B
\begin{equation}\label{jeans0}
    g \gtrsim \sqrt{\frac{T_\nu^2 m_\phi^2}{m_\nu n_\nu}}.
\end{equation}
Since the constraints on $g$ are weak, this condition is easily satisfied across the parameter space.

\begin{figure}
\includegraphics[width=\columnwidth]{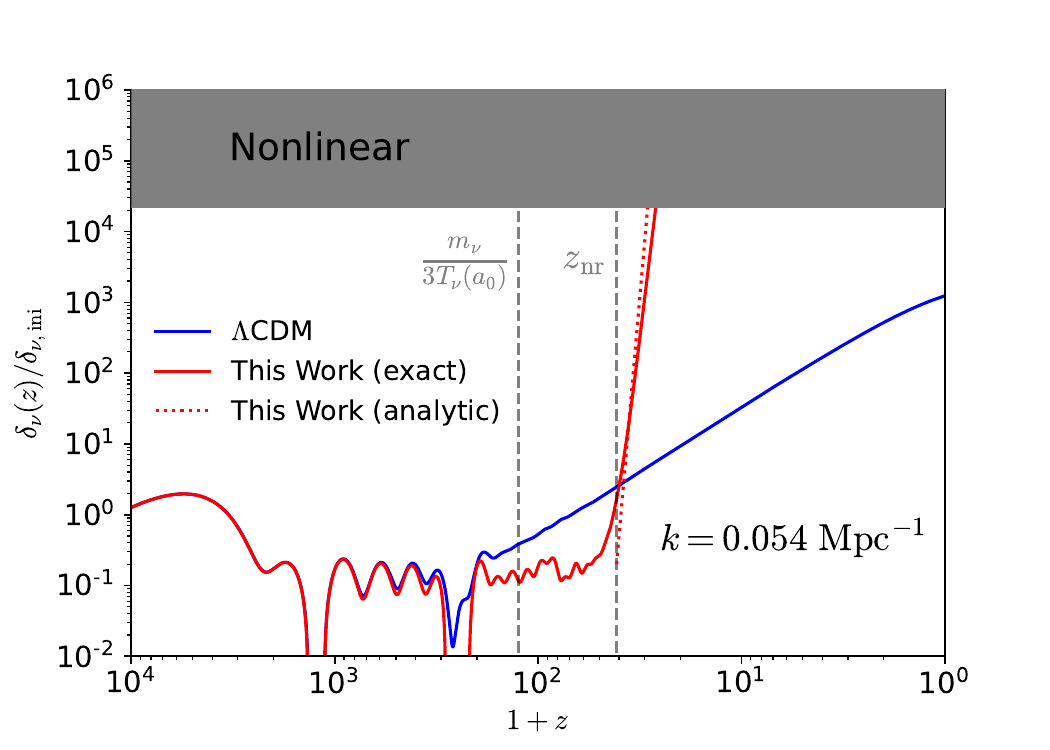}
\caption{The C$\nu$B density perturbation evolution as a function of redshift for wavenumber $k_\phi(a_{\text{NL}}) \approx  0.054 \text{ Mpc$^{-1}$}$. The model parameters are set to $g =10^{-26}$ and $m_\phi = 10^{-29} \text{ eV}$. The vertical axis shows the growth of density perturbations relative to their initial value, with the grey shaded region indicating the nonlinear regime where $\delta_\nu > 1$. The solid lines are obtained by numerically solving the (modified) Boltzmann equation for the C$\nu$B, while the dotted line represents the analytical approximation in Eq.~\eqref{anagrowth}. In the $\Lambda$CDM universe (blue), the C$\nu$B perturbation remains linear at all redshifts. In contrast, the C$\nu$B perturbation in our model (red) grows rapidly after the neutrinos turn non-relativistic at $z_{\text{nr}}\approx 40$ and becomes nonlinear at $z_{\text{NL}} \approx 30$ for the chosen model parameters here.} \label{fig1}
\end{figure}

We now quantitatively study the linear perturbation evolution of the non-relativistic C$\nu$B in the presence of long-range forces. The most straightforward approach is to numerically calculate the modified Boltzmann equation. The exact equations of motion for linear perturbations can be found in Appendix.~\ref{sec:appendix}, which applies for both the relativistic and non-relativistic regimes. We implement these equations of motion into the publicly available code \texttt{CLASS} \cite{Blas:2011rf,Lesgourgues:2011re}. The numerical solutions of the density perturbation evolution of the C$\nu$B is illustrated in Fig.~\ref{fig1}, where the rapid perturbation growth is due to the Jeans perturbation instability.

The perturbation instability occurs when the neutrino becomes non-relativistic. In the non-relativistic limit, the evolution of the C$\nu$B can be effectively described using fluid approximations \cite{Shoji:2010hm, Lesgourgues:2011re, Green:2021gdc}. In the fluid approximation, to leading order, the continuity and Euler equations for the non-relativistic C$\nu$B are \cite{Brookfield:2005bz, Bjaelde:2007ki}: (see also Appendix.~\ref{sec:appendix} for derivations)
\begin{equation}\label{neutrino_euler}
\delta_\nu' = \theta_\nu, \quad \theta_\nu' = -aH\theta_\nu + c_s^2 k^2\delta_\nu + k^2\psi  +k^2\frac{g\delta\phi}{m_\nu},
\end{equation}
where primes denote derivatives with respect to conformal time, $k$ is the wavenumber of perturbations and we take $c_s \approx 3T_\nu/m_\nu$ in our analysis. We also define the fluid density perturbation $\delta_\nu \equiv \delta \rho_\nu / \rho_\nu$ (the complete expressions for $\rho_\nu$ and $\delta \rho_\nu$ can be found in Appendix~\ref{sec:appendix}) and the divergence of the fluid velocity $\theta_\nu \equiv \nabla\cdot \Vec{v}_\nu$. These equations of motion are supplemented by Poisson equations for the gravitational potential $\psi$ and the Yukawa potential $\delta\phi$
\begin{align}\label{poission}
     (k/a)^2\delta \phi + m_\phi^2\delta \phi = -g n_\nu \delta_\nu, \nonumber \\
     (k/a)^2\psi  =  -4\pi G (\delta\rho_m +\delta\rho_\nu),
\end{align}
where $\delta \rho_m$ is the energy density perturbation of the rest of the matter, including cold dark matter (CDM) and baryons. However, for the regime of interest, the perturbation evolution of the non-relativistic C$\nu$B is dominated by the Yukawa force, which is much stronger than gravity. The combination of Eqs.~\eqref{neutrino_euler} and \eqref{poission} gives the equation of motion for $\delta_\nu$ under the fluid approximation:
\begin{equation}\label{delta_eom2}
 \Ddot{\delta}_\nu + 2H\Dot{\delta}_\nu = \frac{3}{2}H^2[(1+\frac{\eta^2k^2}{k^2 +k_\phi^2})f_\nu\delta_\nu - \frac{k^2}{k_{\rm{fs}}^2}\delta_\nu +(1-f_\nu)\delta_{m}],
\end{equation}
where we define $\eta \equiv g/(m_\nu\sqrt{4\pi G})$ and use $H^2 = 8\pi G (\rho_{m}+ \rho_\nu)/3$ in the matter-dominated universe. We also define two characteristic wavenumbers
\begin{align}
    k_{\text{fs}} & \equiv \sqrt{\frac{3}{2c_s^2}}aH   \approx 0.04 \text{ $h$Mpc$^{-1}$} \frac{m_\nu}{60\text{ meV}}\frac{a^2H(a)}{H(a_0)}, \\ k_\phi & \equiv a m_\phi \approx 2.4 \text{ $h$Mpc$^{-1}$} \frac{am_\phi}{10^{-29}\text{ eV}},
\end{align}
which are the free-streaming scale of neutrinos  \cite{2020moco.book.....D} and the comoving wavenumber of the Yukawa interaction range.

In the limit $k \rightarrow 0$, the equation of motion for $\delta_\nu$ in Eq.~\eqref{delta_eom2} becomes similar to that of CDM, since both the free-streaming effect and the Yukawa interactions are finite-distance effects, which disappear in the long-wavelength limit. At smaller wavelengths, both effects become important, and the fate of the C$\nu$B depends on their relative magnitude. Ignoring gravitational interactions, the requirement that the growth mode exists in Eq.~\eqref{delta_eom2} yields the criterion for perturbation instability
\begin{equation}\label{jeans}
    g \gtrsim \sqrt{\frac{c_s^2m_\phi^2m_\nu}{n_\nu}},
\end{equation}
which agrees with our rough estimates in Eq.~\eqref{jeans0}.
 Once the above criterion is satisfied, the perturbation of the C$\nu$B undergoes rapid, non-gravitational growth induced by the Yukawa force. The growth mode can be approximated by
  \begin{equation}\label{anagrowth}
    \delta_\nu(a) \approx \delta_\nu(a_i) (\frac{a}{a_i})^{\gamma(k)}
\end{equation}
where
\begin{align}\label{anagrowth_rate}
    \gamma(k) & = \frac{\sqrt{1+24\left (f_\nu+  \frac{f_\nu \eta^2  k^2}{k^2 + k_\phi^2}  - \frac{k^2}{k_{\text{fs}}^2} \right)}-1}{4} \nonumber \\
    &\approx 47 \left( \frac{k^2}{k^2 + k_\phi^2}\right)^{\frac{1}{2}}\left(\frac{g}{10^{-26}}\right).
\end{align}
The approximation in the second line is valid when $\gamma(k) \gtrsim 1$ and Eq.~\eqref{jeans} holds, which covers the parameter space we are interested in. The wavenumber dependence of $\gamma(k)$ can be understood as follows: the dominant perturbation growth appears at $k \gtrsim k_\phi$ where the interaction range matches with the wavelength.  For $k \lesssim k_\phi$, the interaction range is smaller than the wavelength, resulting in suppression of $\gamma(k)$. At small scales, for $k \gtrsim \sqrt{\eta^2 f_\nu} k_{\text{fs}}$, the growth mode disappears since the Jeans criterion is not satisfied at that scale. The magnitude of $\gamma(k)$ will determine the growth rate of perturbation. When $\gamma(k) \gtrsim 10$ or $g \gtrsim 10^{-26}$, the long-range forces between neutrinos are much stronger than gravity, such that the perturbation growth at $k\sim k_\phi$ is so rapid that a perturbation growth of order $\mathcal{O}\left(10^5\right)$ occurs within a Hubble time. As a result, the perturbation of the C$\nu$B becomes nonlinear almost immediately. For the intermediate regime where $10 \gtrsim \gamma(k) \gtrsim 1$, the perturbation growth lasts for a substantial period of time. But as the universe expands, the perturbation growth will eventually stop when the Jeans criterion in Eq.~\eqref{jeans} is no longer valid. At this point, if $\delta_\nu \lesssim 1$, the C$\nu$B perturbations are enhanced but still insufficient to become nonlinear, and there is no formation of bound states. Finally, for $\gamma(k) \lesssim 1$, the perturbation growth is negligible, and the C$\nu$B will remain smooth. Here, $\gamma(k_\phi) \lesssim 1$ is a similar requirement for $g$ compared to Eq.~\eqref{criterion2}.

Prior to $z_{\text{nr}}$, the C$\nu$B are relativistic and the effective mass of the neutrino is suppressed by $\phi_0$. In this regime, the rapid perturbation growth is absent because the Yukawa interaction between neutrinos is Lorentz suppressed in the relativistic limit. On the other hand, since Eq.~\eqref{criterion1} is a more stringent requirement than Eq.~\eqref{jeans} at $z_{\text{nr}}$, the Jeans criterion generically applies once the neutrinos become non-relativistic. For this reason, we set the initial time of perturbation growth in Eq.~\eqref{anagrowth} at $a_i \approx a_{\text{nr}}$. The transition from stable perturbation evolution in the relativistic limit to rapid perturbation growth in the non-relativistic limit is clearly reflected in Fig.~\ref{fig1}. Compared with the $\Lambda$CDM case, the perturbation evolution of the C$\nu$B in our model is almost identical when they are both relativistic. After $z\sim 100$ the C$\nu$B in $\Lambda$CDM model becomes non-relativistic and undergoes perturbation growth. However, the density perturbation of the C$\nu$B in our model continues to be suppressed because the effective neutrino mass is suppressed. Eventually, at $z_{\text{nr}}$, the C$\nu$B in our model becomes non-relativistic and the rapid perturbation growth shows up, shortly after which the perturbation of the C$\nu$B becomes nonlinear. In Fig.~\ref{fig1}, we also plot the analytical prediction of the perturbation growth in Eq.~\eqref{anagrowth}, which is able to approximate the perturbation growth in the non-relativistic C$\nu$B.

\subsection{Nonlinear structure formation}

The rapid perturbation growth of the C$\nu$B induced by long-range forces eventually leads to the formation of nonlinear neutrino bound states. Although simulations are required to further evolve the system from the first principles, it is still possible to study the formation of bound states by extrapolating results from linear perturbation theory. For example, in the $\Lambda$CDM universe, the Press-Schechter theory and its variants are able to give predictions of the formation of dark matter halos based on perturbation theories \cite{Press:1973iz}. The Press-Schechter theory assumes that a structure collapses into a halo once the matter density perturbation in real space $\delta_m(x)$ exceeds the collapse threshold $\delta_{\text{cr}} \sim 1$. Therefore, the number density of the halos will depend on how much fraction of space exceeds the collapse threshold. This can be calculated based on the fact that the fluctuation of matter density follows a random Gaussian distribution with initial perturbations $\braket{\delta_m^2}^{1/2}_{\text{ini}} \sim 10^{-4}$, which grow over time \cite{2020moco.book.....D}. In the Press-Schechter theory, although the probability that a patch of CDM has collapsed is initially exponentially suppressed, as the perturbation grows to $\delta_m \gtrsim \delta_{\text{cr}}$, the probability becomes close to unity. Physically, this means that when the density fluctuations are large, the matter density field should collapse into halos.

The formation of nonlinear structures induced by the Yukawa force has two main differences compared to the standard structure formation induced by gravity. First, since the Yukawa force is much stronger than gravity, the time scale for the perturbation growth is much shorter than the Hubble time, so that the perturbations grow to exceed $\delta_\nu \gtrsim 1$ very quickly. Upon reaching this threshold, we expect an order one fraction of the C$\nu$B to form nonlinear bound states. Secondly, the finite range of the Yukawa force causes the perturbation growth to be cut off at $k \lesssim k_\phi$. This results in nonlinear structures being preferentially formed at the wavenumber $k_\phi$. Although these two features can be observed by applying the Press-Schechter formalism to $\delta_\nu$, it has not been rigorously checked whether the Press-Schechter theory is applicable to non-gravitational collapses. However, previous simulation studies on Yukawa force induced structure formation indeed show a peaked halo mass function \cite{Domenech:2023afs}.

\begin{figure}
\includegraphics[width=\columnwidth]{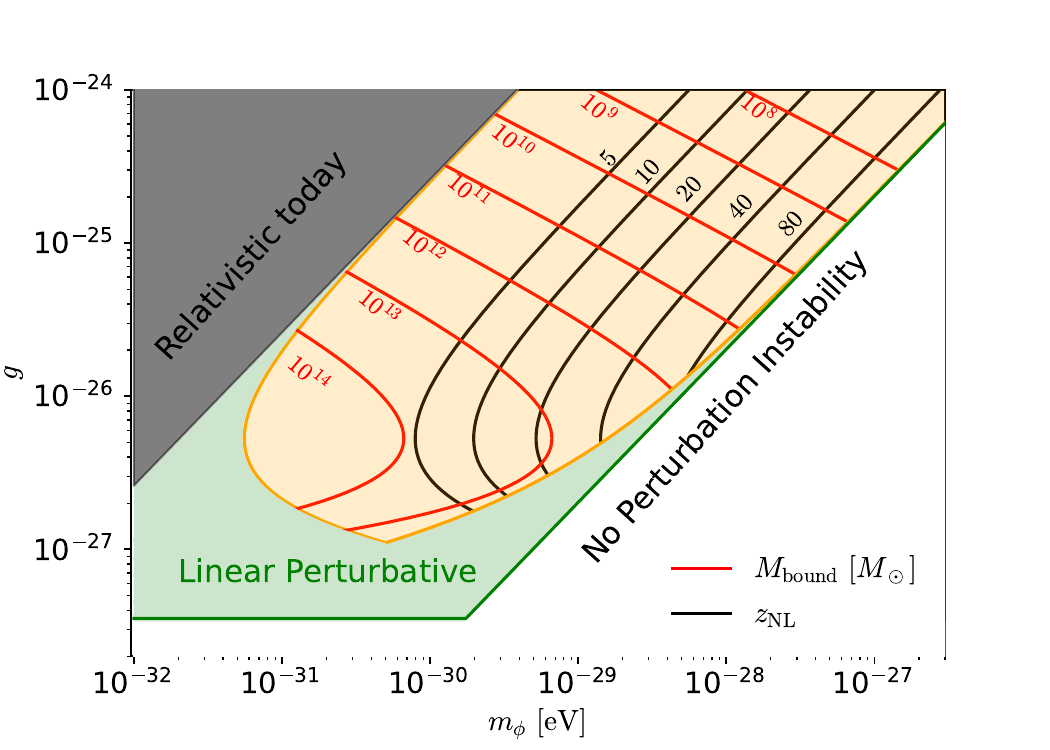}
\caption{Overview of the parameter space. In the area shaded in grey, the neutrino mass is suppressed today. Above the green solid line, the long-range forces between neutrinos can induce perturbation instability. But only in the orange shaded regime, the perturbation growth is sufficient to form nonlinear neutrino bound states. In the orange regime, we plot the estimated mass of the neutrino bound states $M_{\text{bound}} \ [M_\odot]$ and the formation redshift $z_{\text{NL}}$. In the green shaded regime, the C$\nu$B perturbations remain linear, but the background evolution of $\phi_0$ can still result in the suppression of neutrino mass. Below the green solid line, both the background and perturbation evolutions of $\phi$ do not significantly affect the C$\nu$B.}\label{mass_relation}
\end{figure}

Based on the above discussions, we assume that in the presence of long-range forces, the C$\nu$B forms nonlinear bound states when $\delta_\nu(k_\phi)$ becomes larger than unity. Upon formation, most of the C$\nu$B within the interaction range $m_\phi^{-1}$ collapses into a bound state. We estimate the mass of the bound states to be \footnote{Here we do not take into account the finite $\phi_0$ upon formation. There are two effects: first, $\phi_0$ modifies the neutrino mass; second, $\phi_0$ also contains energy density. We find at most an order one change in the total energy density of the $\nu-\phi$ system when these effects are considered, which we will not include in our analysis.}
\begin{equation}\label{eq:mass}
    M_{\text{bound}} \sim \frac{4\pi m_\nu n_{\nu,0}(a_{\text{NL}})}{3 m_\phi^3},
\end{equation}
where $a_{\text{NL}}$ denotes the scale factor at which $\delta_\nu(k_\phi)$ becomes nonlinear, and we denote $z_\text{NL} = 1/a_{\text{NL}} - 1$. As discussed in Sec.~\ref{sec:lpt}, in the regime where $\gamma(k)\gtrsim 10$ or $g \gtrsim 10^{-26}$, $\delta_\nu$ exceeds unity within one Hubble time such that the nonlinear bound states formation follows immediately after the C$\nu$B becomes non-relativistic at $a_{\text{NL}} \approx a_{\text{nr}}$. In this regime, from Eq.~\eqref{eq:nr}, the background number density of neutrinos at $a_{\text{NL}}$ is approximately $n_{\nu,0}(a_{\text{NL}}) \approx m_\phi^2 m_\nu/g^2$, and Eq.~\eqref{eq:mass} becomes
\begin{align}\label{mass_powerlaw}
    M_{\text{bound}} \sim 1.4\times 10^{13}\ M_{\odot}\left(\frac{g}{10^{-26}}\right)^{-2}\left(\frac{m_\phi}{10^{-29}\text{ eV}}\right)^{-1}.
\end{align}
More generally, we estimate $a_{\text{NL}}$ based on Eqs.~\eqref{eq:nr} and \eqref{anagrowth}. In particular, we set the start of perturbation growth at $z_{\text{nr}}$, then apply $\gamma(k_\phi)$ to Eq.~\eqref{anagrowth} to estimate the perturbation growth. We require that before Eq.~\eqref{jeans} becomes invalid and before today ($z_{\text{NL}}>0$), the perturbation growth must be at least $10^5$ so that $\delta_\nu(k_\phi) \gtrsim 1$. We plot the regime that satisfies these semi-analytical requirements in the orange-shaded area in Fig.~\ref{mass_relation}. Within the regime where neutrino bound states can form, we also plot the estimated $a_{\text{NL}}$ and $M_{\text{bound}}$ based on Eq.~\eqref{eq:mass}. When numerically solving the evolution of $\delta_\nu$ using the modified \texttt{CLASS} code, we find that these semi-analytical estimations of $z_{\text{NL}}$ are consistent with the results of numerical calculations. At $m_\phi \gtrsim 10^{-28} \text{ eV}$, the code breaks down due to numerical problems with highly oscillatory functions. After that, analytical estimations become necessary.
In Fig.~\ref{mass_relation}, the deviation from the power-law estimation in Eq.~\eqref{mass_powerlaw} appears at $g\lesssim10^{-26}$ or $\gamma(k) \lesssim 10$ because the formation of nonlinear bound states is not instantaneous. In the other direction, as $m_\phi$ and $g$ increase, the interaction range decreases, leading to the formation of smaller nonlinear structures.

In the standard theory of structure formation, an overdense patch of CDM collapses into a virialized halo with a radius $\mathcal{O}(1)$ times smaller than the turn-around radius. For the non-relativistic C$\nu$B with the momentum distribution given in Eq.~\eqref{eq:distribution}, the momentum space is not close to being filled, enabling further contraction during the collapse. In Eq.~\eqref{eq:mass}, the turn-around radius is assumed to be $m_\phi^{-1}$; therefore, the radius of the bound state is at most $R_{\text{bound}} < m_\phi^{-1}$. However, these neutrino bound states differ from CDM halos in several ways.  For example, the sourced $\phi$ can back-react on the neutrinos as the bound state contracts, and the constituents in the neutrino bound states are near-relativistic to start with. Therefore, determining the mass-radius relation and the profile of the neutrino bound states is non-trivial. In this regard, many studies have looked for stable bound state solutions in similar systems \cite{Stephenson:1996qj,Smirnov:2022sfo,Brouzakis:2007aq}.
Contracting to a denser object will increase the gravitational attraction to surrounding matter, thus enhancing the signals discussed in Sec.~\ref{sec:cos}. In this work, we will conservatively assume that the radii of the neutrino bound states are close to the maximum radius
\begin{equation}\label{eq:radius}
    R_{\text{bound}} \sim m_\phi^{-1} \approx 640 \text{ kpc} \left(\frac{m_\phi}{10^{-29}\text{ eV}}\right)^{-1}.
\end{equation}
Then the energy density of these bound states is just the background energy density of the C$\nu$B at $z_{\text{NL}}$, which is 
\begin{equation}\label{eq:density}
    \rho_{\text{bound}} \sim m_\nu n_{\nu,0} (z_{\text{NL}}) \approx 6.7 \times 10^{-3} \left(\frac{1+z_{\text{NL}}}{100}\right)^{3} \text{ GeV}/\text{cm}^3.
\end{equation}

It is speculated that these bound states form with a small net velocity \cite{Afshordi:2005ym, Esteban_2021}. The C$\nu$B prior to the collapse has only negligible bulk velocity. During the collapse, the infall of neutrinos from different directions also cancels the net velocity. Only rarely can a highly asymmetric collapse contribute significantly to the net velocity of the bound states. However, upon formation, these bound states occupy an order one fraction of the volume of the universe, such that even if one bound state gains a large net velocity, it can dissipate its net velocity into the surrounding bound states through coherent scatterings. For this reason, within a Hubble time after the formation, the bound states should not move a significant distance compared to their typical separation, which is of the order of their radius. Therefore, in the subsequent analysis we will ignore their net velocity.

Once the neutrino bound states have formed and virialized, for the regime of parameter space we are interested in, the bound states do not evolve further and remain stable. The energy loss due to $\phi$ emission from neutrino scattering and annihilation is negligible throughout. Additionally, the relaxation time scale of the bound states is much longer than the age of the universe \cite{2008gady.book.....B}, hence the escape of neutrinos can also be ignored. However, prior to the virialization of the neutrino bound states, their density distribution can be asymmetric, and coherent emission of $\phi$ and gravitational waves can be present during the collapse and virialization. This is similar to the gravitational wave production induced by the nonlinear structure formation discussed in Refs.~\cite{Flores:2022uzt,Fernandez:2023ddy,Eggemeier:2022gyo,Dalianis:2020gup,Jedamzik_2010}, but occurs in the late time universe. For the parameter space shown in Fig.~\ref{fig:exclusion1}, the nonlinear neutrino structures are too large, and the gravitational waves produced have wavelengths that are too long to be observed by current gravitational wave observatories \cite{Schmitz:2020syl}. Whether the formation of smaller nonlinear neutrino structures can produce observable gravitational wave signals would be an interesting subject for future studies.

\section{Signals}\label{sec:cos}

In the previous section, we demonstrated that long-range forces between neutrinos can induce rapid, non-gravitational perturbation growth in the C$\nu$B. This can lead to the formation of nonlinear bound states, and we have estimated their essential properties including mass, radius, and formation redshifts. In the standard cosmological model, the formation of massive halos at high redshift is very suppressed. If massive nonlinear neutrino structures can form at high redshift, they should have significant impacts on structure formation in the universe, which we quantify in this section.

\subsection{Imprints on structure formation}

As the bound states of the C$\nu$B interact with CDM and baryons through gravity, they contribute additional perturbations to the matter density field. On the large scale, the matter density perturbation is smooth. The evolution of the combined density field of neutrino bound states and CDM can be described by linear perturbation theory. Therefore the total linear matter power spectrum observed today is a combination of CDM perturbations (labeled ``ad'') and neutrino bound states induced perturbations (labeled ``iso'', uncorrelated with CDM)
\begin{align}\label{mps_tot}
    P_{\text{m}}(a,k) = P_{\text{ad}}(a,k) +P_{\text{iso}}(a,k). 
\end{align}
A discrete distribution of matter will contribute a minimum amount of matter power spectrum from Poisson fluctuations even if they are uncorrelated in real space. This has been studied in the PBH \cite{Afshordi:2003zb,Murgia:2019duy,Inman:2019wvr,Delos:2024poq,Carr:2018rid,Liu:2022bvr,Liu:2022okz,Liu:2023pvq,Zhang:2023hyn,Zhang:2024ytf} and MACHO \cite{Bai:2020jfm,Croon:2024rmw,Croon:2024jhd,Irsic:2019iff,Chang:2024fol,Amin:2022nlh} models. Based on these, we estimate the perturbations induced by the nonlinear bound states of neutrinos to be
\begin{align}\label{Poisson}
    P_{\text{iso}}(a,k) \approx \frac{f_\nu^2 D_+(a,a_{\text{NL}})^2 }{\Bar{n}_{\text{bound}}},
\end{align}
where $\Bar{n}_{\text{bound}} = m_\nu n_{\nu,0}(a_0)/M_{\text{bound}}$ is the average comoving number density of the neutrino bound states and $D_+(a,a_{\text{NL}})$ is the growth factor of perturbations from $a_{\text{NL}}$ to $a$. Since only the perturbation growth of sub-horizon modes in matter-dominating universe is relevant, the growth factor is approximately $D_+(a,a_{\text{NL}}) \approx a/a_{\text{NL}}$.

\begin{figure}
\includegraphics[width=\columnwidth]{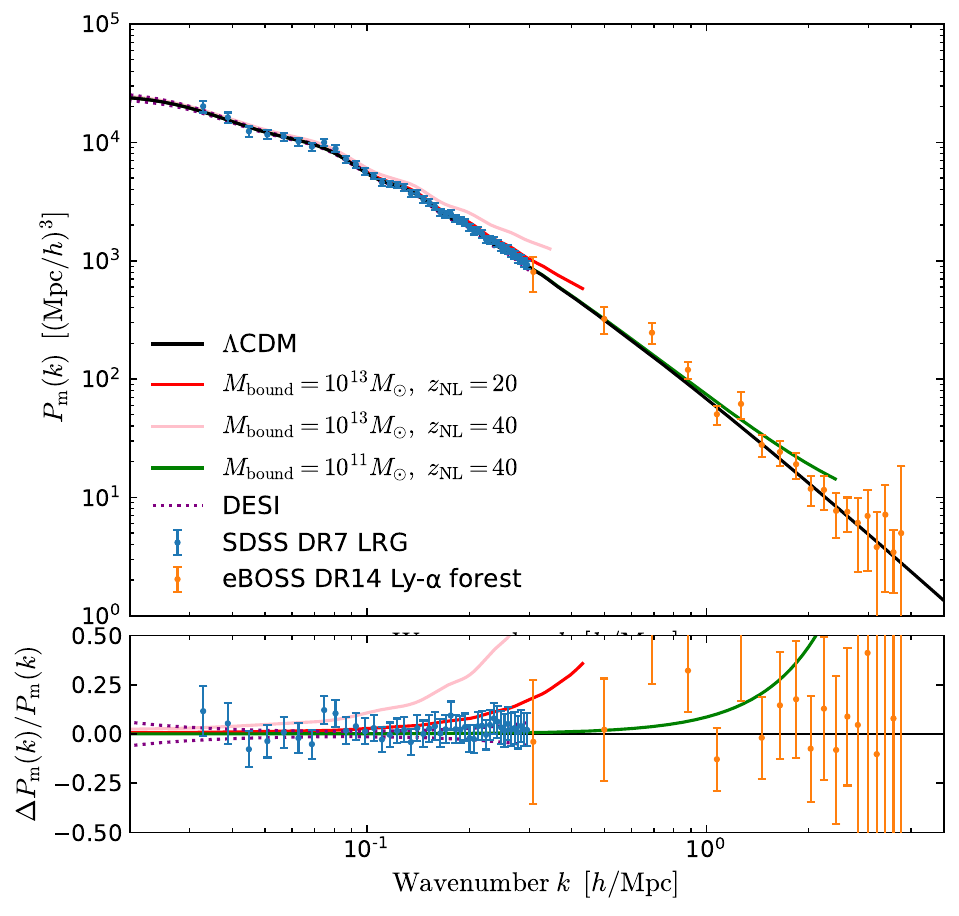}
\caption{Illustration of the impact of the formation of neutrino bound states on the linear matter power spectrum. Here the black curve is the expected linear matter power spectrum from the fiducial cosmology at $z=0$. The colored solid curves are the matter power spectrum evaluated based on Eq.~\eqref{mps_tot} at the corresponding $z_{\text{obs}}$ and then extrapolated to $z=0$. The curves are cut off at $ k_{\text{cut}}(z_{\text{obs}})$, where nonlinear effects become important.}\label{result}
\end{figure}

The above linear perturbation description of the matter power spectrum breaks down at a small scale $k_{\text{cut}}$ where the matter density perturbation becomes nonlinear around the neutrino bound states\footnote{Eq.~\eqref{Poisson} also becomes invalid at small scales where adiabatic matter perturbations become nonlinear. However, this is irrelevant for our purposes since we only discuss the modifications to the linear regime of the matter power spectrum.
}. After which, the CDM and baryons collapse and virialize into halos clothed surrounded the neutrino bound states, halting further perturbation growth. We expect Eq.~\eqref{Poisson} to hold up to the wavenumber $k_{\text{cut}}$ that corresponds to the turn-around radius $r_{\text{ta}}$ of the secondary collapse.
Studies on the spherical collapse model \cite{Fillmore:1984wk, Bertschinger:1985pd, Mack:2006gz, Vogelsberger:2009bn, Ludlow:2010sy, Bringmann:2011ut} predict that the total mass of the secondary collapsed halo (minihalo) grows linearly with the scale factor and the turn-around radius is given by
$4\pi r_{\text{ta}}^3\rho_{m}/3 = M_{\text{bound}}\frac{a}{a_{\text{NL}}}$. Based on these, we estimate that Eq.~\eqref{Poisson} starts to break down due to nonlinear effects at
\begin{align}\label{cutoff}
    k \gtrsim k_{\text{cut}} \equiv \frac{2\pi a}{2r_{\text{ta}}} = \pi \left(\frac{4\pi \bar{n}_{\text{bound}}}{3f_\nu}\right)^{1/3} \left(\frac{a}{a_{\text{NL}}}\right)^{-1/3} \nonumber \\
    \approx 1.2 \text{ $h$Mpc$^{-1}$} \left( \frac{M_{\text{bound}}}{10^{13} M_\odot}  \right) ^{-1/3} \left(\frac{a}{a_{\text{NL}}}\right)^{-1/3}.
\end{align}
 In the following analysis, we will conservatively only use Eq.~\eqref{Poisson} in the wavenumber range $k < k_{\text{cut}}$ and set $P_{\text{iso}}(k>k_{\text{cut}}) =0$\footnote{The choice of nonlinear scale as cutoff was also used in Refs.~\cite{Liu:2022bvr,Gouttenoire:2023nzr}. The formula is different because we have taken into account the fact that as perturbations evolve, the $k_{\text{cut}}$ will gradually decrease corresponding to the collapse of larger structures.}.

\begin{figure}
\includegraphics[width=\columnwidth]{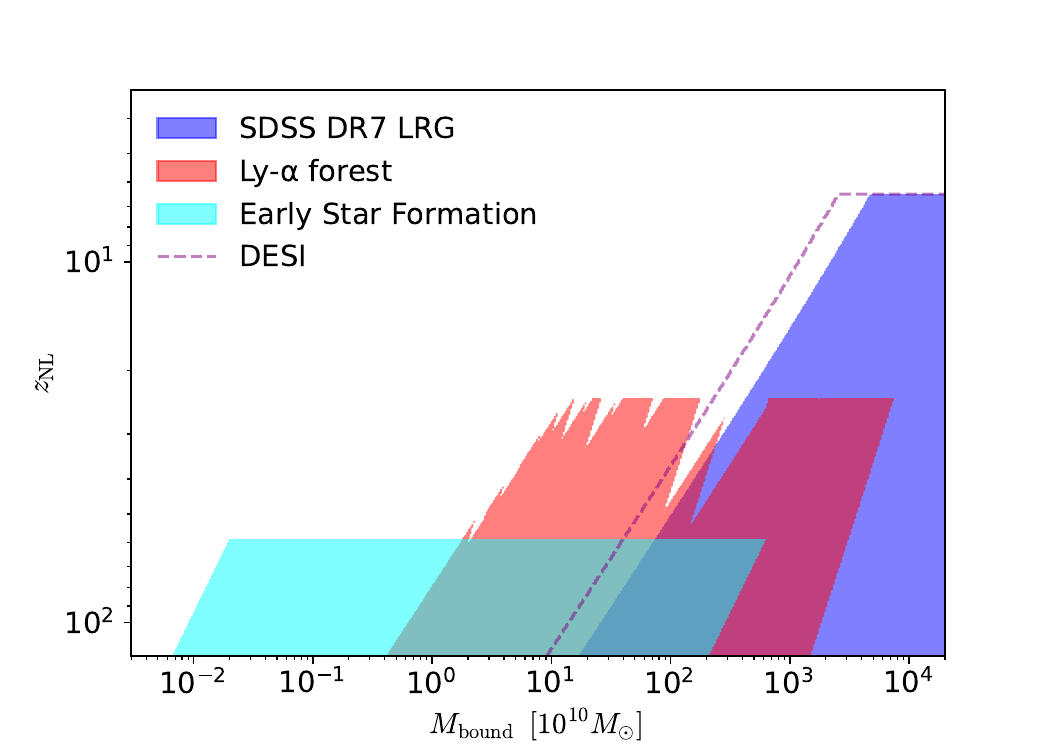}
\includegraphics[width=\columnwidth]{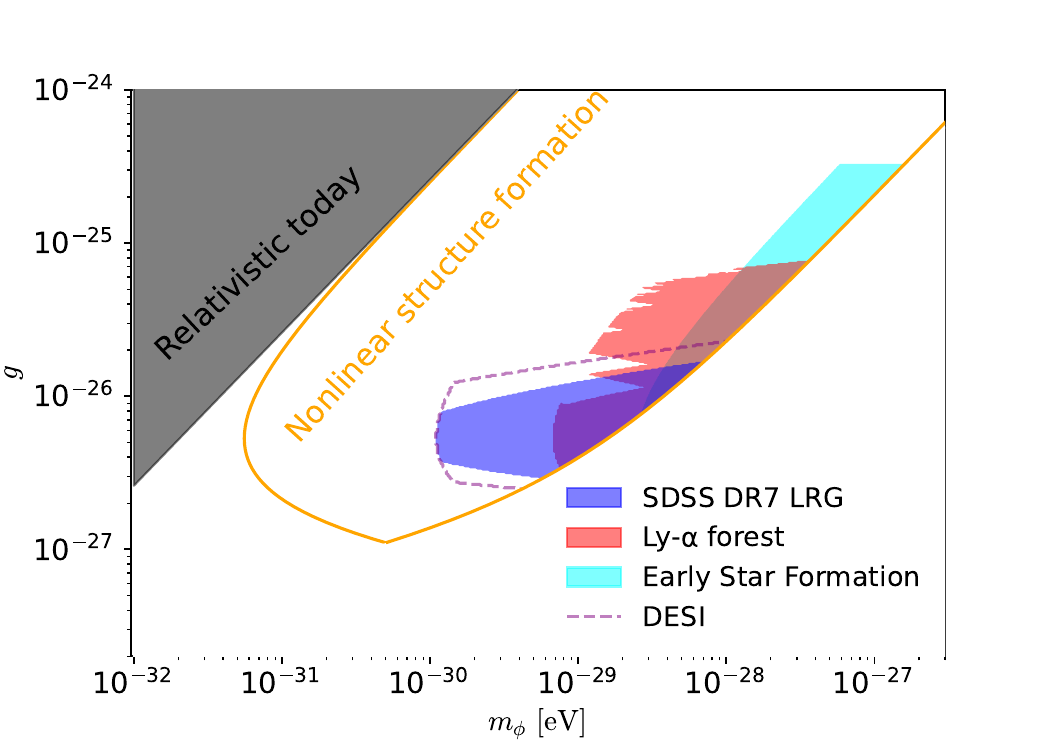}
\caption{Constraints on the formation of neutrino bound states. \textit{Top:} Constraints on the parameter space $(M_{\text{bound}},z_{\text{NL}})$. The shaded red (blue) area is ruled out by the Ly$\alpha$ forest (galaxy clustering) measurements of the linear matter power spectrum. The dashed purple line represents the estimated sensitivity from DESI galaxy surveys. In the shaded cyan areas, star formation at $z\gtrsim 7 $ is expected to be triggered by neutrino bound states, which is constrained by the reionization history measured from CMB observations. \textit{Bottom: }Similar to the top plot, but mapping the constraints on $(M_{\text{bound}},a_{\text{NL}})$ to the model parameter space $(g, m_\phi)$.}\label{fig:exclusion1}
\end{figure}

So far, we have treated the neutrino bound states as point particles. In our analysis, we only consider the wavenumbers at $k<k_{\text{cut}}$, corresponding to a spatial resolution of $r_{\text{ta}}$. If the radii of these bound states are much smaller than $r_{\text{ta}}$, their finite sizes become irrelevant to our analysis, allowing them to be treated as point particles. Therefore, in our analysis, we require that the radii of these bound states satisfy $2R_{\text{bound}} < r_{\text{ta}}$ at the time of observation, which corresponds to
\begin{align}\label{eq:finite_size}
    a_{\text{obs}} > 2^{\frac{3}{4}}f_\nu^{-1/4}a_{\text{NL}} \approx 6.5a_{\text{NL}}.
\end{align}
Physically, this condition also ensures that neutrino bound states are sufficiently dense to induce the collapse of the surrounding matter at $a_{\text{obs}}$. In the upper panel of Fig.~\ref{fig:exclusion1}, Eq.~\eqref{eq:finite_size} results in the horizontal cutoff of the constraints imposed by the Ly$\alpha$ forest and galaxy clustering observations.

To set the constraints on long-range forces between neutrinos, we use the 3D linear matter power spectrum reconstructed from the Ly$\alpha$ forest 1D power spectrum (Ly$\alpha$) and from the halo power spectrum for galaxy clustering (LRG) in Ref.~\cite{Chabanier:2019eai} (see also Refs.~\cite{DES:2017qwj,Planck:2018nkj,Sabti:2021unj,Gilman:2021gkj,Esteban:2023xpk,LSSTScience:2009jmu,EuclidTheoryWorkingGroup:2012gxx} for other matter power spectrum measurements). These data sets are measured at different redshifts, $z_{\text{obs}}\sim 0.35$ for LRG, $z_{\text{obs}} \sim 2.2$ to $4.6$ for Ly$\alpha$. The matter power spectra are evaluated at the observation redshift and extrapolated to $z=0$ based on linear perturbation theory. The combination of these two data sets covers the wavenumber range from $k \approx 2\times 10^{-2}\  h\text{Mpc}^{-1}$ to $k \approx 2 \  h\text{Mpc}^{-1}$. Similarly to Refs.~\cite{Green:2021xzn,Green:2021gdc}, we also estimate the sensitivity of the Dark Energy Spectroscopic Instrument (DESI) based on the volume of the survey (which sets the cosmic variance at large scales) and the number of galaxies (which raises the shot noise at small scales) given in Ref.~\cite{Font-Ribera:2013rwa}.

We plot the measured linear matter power spectrum in Fig.~\ref{result}. To illustrate how constraints vary, we also show the modified  spectrum for different phenomenological parameters $(z_{\text{NL}}, M_{\text{bound}})$. In terms of the model parameters, the red (green) curve corresponds to $g \approx 8.2\times 10^{-27}$, $m_\phi \approx 5.6 \times 10^{-30} \text{ eV}$ ($g \approx 4.6\times 10^{-26}$, $m_\phi \approx 5.1 \times 10^{-29} \text{ eV}$). The phenomenological parameters for the pink curve are not viable for the default neutrino mass but become accessible for larger neutrino masses. For proper comparison with observations, we set $z_{\text{obs}} = 0.35$ for the red and pink curves and $z_{\text{obs}} = 3$ for the green curve.

We then obtain the exclusion limit on long-range forces between neutrinos in the following way. We first evaluate the linear matter power spectrum $P_{\text{ad}}(k)$ in the standard $\Lambda$CDM model with the fiducial cosmological parameters from Planck \cite{Planck:2018nkj}. Then we turn on  long-range forces between neutrinos such that the C$\nu$B will form nonlinear bound states characterized by $(z_{\text{NL}}, M_{\text{bound}})$. The corresponding radius $R_{\text{bound}}$ can be solved from Eq.\eqref{eq:mass} and \eqref{eq:radius}. The modified matter power spectrum is evaluated from Eqs.~\eqref{mps_tot}, \eqref{Poisson} and \eqref{cutoff} at the observation redshift $z_{\text{obs}}$ (we set $z_{\text{obs}} = 0.35$ for LRG and $z_{\text{obs}} = 3$ for Ly$\alpha$ in our analysis) and extrapolated to $z= 0$. The imprints of the neutrino bound states have a different wavenumber dependence compared to the standard matter power spectrum, which can hardly be mimicked by changing other cosmological parameters. So we expect the constraints on $(z_{\text{NL}}, M_{\text{bound}})$ to be not in degenerate with the rest of cosmological parameters, and we will keep them fixed to the fiducial values in our analysis. The $\chi^2$ function is defined as
\begin{align}
    \chi^2(z_{\text{NL}}, M_{\text{bound}})  = \sum_{i=1}^N   (\frac{P_{\text{obs}}(k_i) - P_m(z_{\text{NL}}, M_{\text{bound}},k_i)}{\sigma_{P(k_i)}})^2,
\end{align}
where $P_{\text{obs}}(k_i)$ and $\sigma_{P(k_i)}$ are the reconstructed 3D linear matter power spectrum at wavenumber $k_i$ and the associated error. We first compute $\chi^2_0$ from fiducial cosmology. Then for each $M_{\text{bound}}$, we vary $z_{\text{NL}}$ and evaluate $\chi^2(z_{\text{NL}}, M_{\text{bound}})$. We expect that the difference $\chi^2(z_{\text{NL}}, M_{\text{bound}})-\chi^2_0$ follows a half $\chi^2$ distribution \cite{Cowan:2010js}. $\chi^2(z_{\text{NL}}, M_{\text{bound}}) - \chi^2_0 > \chi^2_{95\%}$ sets the $95\%$ confidence limit on $(z_{\text{NL}}, M_{\text{bound}})$, where $\chi^2_{95\%} \approx 2.71$.

The constraints on $(z_{\text{NL}}, M_{\text{bound}})$ can be mapped to the model parameters $(g, m_\phi)$ based on the analysis in Sec.~\ref{sec:lpt}, using Eqs.~\eqref{eq:nr}, \eqref{anagrowth_rate} and \eqref{eq:mass}. We illustrate both the constraints on $(z_{\text{NL}}, M_{\text{bound}})$ and $(g, m_\phi)$ in Fig.~\ref{fig:exclusion1}.

In this section, our analytical model is limited to the regime where matter perturbations are linear and neutrino bound states are small. To investigate beyond this regime, simulations may be necessary. We highlight two effects that are not included in the analytical model. First, in the parameter space where Eq.~\eqref{eq:finite_size} is invalid, we do not impose any constraints. However, diffuse bound states can still enhance structure formation through the gravity from their overdensities, which might be statistically distinguishable in galaxy clustering data. Second, since these bound states form through highly nonlinear evolution, their density fields are expected to be very non-Gaussian. For non-Gaussian stochastic fields, the power spectrum used here is insufficient to fully describe the system \cite{Cheng:2020qbx}. We expect future simulation works to be able to explore a much broader parameter space when these effects are included.

\subsection{Baryonic structure formation}

In the standard cosmological model, massive CDM halos will gravitationally trap baryons. As gas falls into these dark matter halos, the baryons undergo radiative cooling, leading to runaway gravitational collapse. Subsequently, these baryon clouds fragment and eventually collapse into stars due to the Jeans instability. To allow the whole mechanism to work, a key requirement is efficient cooling of the baryonic gas. This efficient cooling is typically expected to occur when the Rees-Ostriker-Silk (ROS) criterion is satisfied \cite{Silk:1977wz, Bromm:2003vv,Bromm:2013iya,Sullivan:2018szg,2010gfe..book.....M,2010hdfs.book.....L}, which requires the free-fall time scale $t_{\text{ff}} \sim  (R_{\text{bound}}^3/GM_{\text{bound}})^{1/2}$ to be longer than the gas cooling time scale $t_{\text{cool}} \sim 3n T_{\text{vir}}/2 \Lambda_{\text{cool}}$, where $n$ is the gas number density, $T_{\text{vir}}$ is the (gravitational) virial temperature of the bound state, and $\Lambda_{\text{cool}}$ is the gas volume cooling rate.

 Based on Ref.~\cite{Barkana:2000fd}, the gas cooling rate $\Lambda_{\text{cool}}$ is kinematically suppressed at low temperatures, but turns on very steeply if the temperature exceeds 
\begin{equation}\label{eq:ros}
    T_{\text{vir}} \sim \frac{GM_{\text{bound}}m_H}{R_{\text{bound}}} \gtrsim 10^4 \text{ K},
\end{equation}
which corresponds to the activation of radiative cooling from collisional excitations of atomic hydrogens. Here $m_H$ is the mass of the hydrogen atom. From Ref.~\cite{Barkana:2000fd}, the gas volume cooling rate is $\Lambda_{\text{cool}}/n^2 \sim 10^{-22}$ erg$\,$cm$^3/$s for $T_{\text{vir}} \gtrsim 10^4 \text{ K}$, which is capable of cooling baryonic gas very efficiently.  For example, at $z \sim 7 $ and $T_{\text{vir}} \sim 10^4 \text{ K}$, we estimate that the cooling time scale is $t_{\text{cool}} \sim \mathcal{O}( \text{Myr})$, which is much smaller than the corresponding Hubble time $1/H(z=7) \approx 1.2\  \text{Gyr}$. For $z \sim 7$, when substituting the $M_{\text{bound}}$ and $R_{\text{bound}}$ of neutrino bound states into Eq.~\eqref{eq:ros}, we find that Eq.~\eqref{eq:ros}  requires a minimum mass of $M_{\text{bound}} \gtrsim 10^8 M_{\odot}$ (see also the upper panel of Fig.~\ref{fig:exclusion1}), which is similar to the minimum star-forming halo mass estimated in Refs.~\cite{Barkana:2000fd,Bromm:2013iya,Sullivan:2018szg,Bromm:2003vv,2010hdfs.book.....L}. Furthermore, we will restrict our analysis to $T_{\text{vir}} \lesssim 10^7 \text{ K}$, where the cooling time scale is still much shorter than the Hubble time for the relevant redshift. The cooling rate will also depend on the chemical abundance of the baryonic gas, such as molecular hydrogen fraction and metallicity, but including these will only further enhance the cooling rate and thus also enhance star formation \cite{2010gfe..book.....M}.

In order to capture baryonic gas, it is also necessary to require that the dynamical time scale is shorter than the Hubble time
\begin{align}\label{eq:time}
    t_{\text{ff}} \sim  \left( \frac{R_{\text{bound}}^3}{GM_{\text{bound}}}\right)^{1/2} \lesssim \frac{1}{H}.
\end{align}
This condition sets the minimum energy density to have star formation in bound states. Upon the formation of neutrino bound states, their energy density is given by Eq.~\eqref{eq:density}, and Eq.~\eqref{eq:time} is not satisfied. Then, as the universe expands, this condition can be satisfied at a lower redshift. For Eq.~\eqref{eq:time} to be satisfied at $z\sim 7$ requires $z_{\text{NL}} \gtrsim 60$, which results in the horizontal cutoff in the cyan shaded area in the upper panel of Fig.~\ref{fig:exclusion1}. In addition to the requirements in Eqs.~\eqref{eq:ros}
and \eqref{eq:time}, the (gravitational) virial velocity of the neutrino bound
states must exceed both the net velocity and the thermal velocity of baryons to effectively capture them \cite{Tseliakhovich:2010bj,Gnedin:1997td,Naoz:2012fr,Arvanitaki:2019rax}. However, these requirements are much weaker than those imposed by Eq.~\eqref{eq:ros} for the relevant redshifts.

When both Eqs.~\eqref{eq:ros} and \eqref{eq:time} are satisfied, the neutrino bound states will have the correct time scale for star formation: $t_{\text{cool}} \ll t_{\text{ff}} \lesssim  1/H$. This ensures that the neutrino bound states can gravitationally capture nearby matter, and then radiative cooling can efficiently cool down the captured baryonic gas. Following this, we expect the subsequent gas dynamics to lead to star formation within these neutrino bound states.

The current measurement of the CMB optical depth suggests that reionization occurs at $z\sim 7$ \cite{Planck:2018nkj}. The cosmological transition from neutral hydrogen gas to ionized hydrogen gas is believed to be caused by ionizing radiation produced by stars in the early galaxies \cite{Barkana:2000fd,2010hdfs.book.....L,Yoshida:2003rw,Munoz:2021psm,Cain:2024fbi,Cruz:2024fsv,Munoz:2024fas}. Assuming Population II stars are producing these ionizing photons, it is estimated in Ref.~\cite{2010hdfs.book.....L} that an $\mathcal{O}(1\%)$ fraction of baryons collapsing into galaxies is enough to reionize the universe. Therefore, we expect that in our case, if star formation can take place in these neutrino bound states at $z \gtrsim 7$, since these neutrino bound states can capture an $\mathcal{O}(f_\nu)$ fraction of matter, the reionization history will be significantly altered and thus be tightly constrained by CMB observations. In Fig.~\ref{fig:exclusion1}, we highlight in cyan the parameter space where Eqs.~\eqref{eq:ros} and \eqref{eq:time} are satisfied at $z > 7$. In practice, non-standard ionization photon injections during the cosmic dark ages can be tested with much better accuracy than an $\mathcal{O}(1)$ level if the photon source is known \cite{Slatyer:2016qyl,Capozzi:2023xie,Xu:2024vdn,Sun:2023acy,Qin:2023kkk}. Moreover, non-standard structure formation during the cosmic dark ages can also be probed by future 21-cm line observations \cite{Munoz:2019hjh,deKruijf:2024voc,Jones:2021mrs,Vanzan:2023gui,Hotinli:2021vxg,Flitter:2022pzf,Munoz:2019hjh,Munoz:2016owz,Cole:2019zhu,Short:2022bmm,Driskell:2022pax,Ali-Haimoud:2013hpa}.

Besides altering the reionization history, star formation within neutrino bound states can lead to a wide variety of observables beyond standard galaxy formation. The formation of high-redshift galaxies can be probed by the UV luminosity function of galaxies observed by the Hubble Space Telescope \cite{Sabti:2021unj} or be directly observed by the James Webb Space Telescope \cite{Boylan-Kolchin:2022kae}. Additionally, the profiles of the neutrino bound states can be very different from those of usual CDM halos. Therefore, galaxies hosted by the neutrino bound states can have kinematics and structures distinctive from those hosted by CDM halos, which can be searched for similarly to that in Ref. \cite{Esteban:2023xpk,Kim:2021zzw}. Also, the halo mass function of these neutrino bound states is expected to peak at the mass scale that corresponds to the Yukawa interaction range, which differs from that of the CDM halos.

\section{Discussion and Conclusion}\label{sec:con}

In this work, we explored the consequences for cosmic structures in the presence of new long-range forces between neutrinos. We studied the linear perturbation evolution of the C$\nu$B both numerically and analytically, and found that the non-relativistic C$\nu$B can experience rapid, non-gravitational perturbations growth, causing them to become nonlinear in the late time universe. As a result, nonlinear bound states of neutrinos will form, and we estimated their formation redshift, mass, and radius.

The formation of these nonlinear neutrino bound states can significantly enhance structure formation. We quantified their gravitational influence on the linear matter power spectrum based on the Poisson fluctuations in the density field and found that the current measurements of the linear matter power spectrum already place constraints on long-range forces between neutrinos. We also found that these bound states can satisfy the necessary requirements for star formation. Early star formation inside the neutrino bound states can modify the reionization history, which is constrained by CMB observations.

Throughout this work, we considered $\phi$ coupling to only one neutrino species. If $\phi$ also couples to other neutrinos species, the perturbation instability still generically occurs, as long as $\phi$ is dominantly coupled to non-relativistic neutrinos. To take an extreme example, suppose there are two neutrinos with the same mass but opposite coupling to $\phi$. Since the opposite couplings cancel each other, the evolution of $\phi_0$ becomes decoupled from the background neutrinos.  Nevertheless, the perturbation instability still arises when neutrinos become non-relativistic, as the Yukawa force will tend to segregate the density fields of the two different neutrinos. On the other hand, if $\phi$ is dominantly coupled to relativistic neutrinos, it will tend to suppress the perturbation instability by increasing $m_{\phi,\text{th}}$. Since this work only considered the gravitational effects of these bound states, we expect the signals to not strongly depend on how neutrinos couple to $\phi$ as long as the resulting nonlinear structures formed through the perturbation instability are similar.

Although we have mainly focused on the phenomena associated with very massive neutrino bound states ($M_{\text{bound}}\gtrsim10^8 M_{\odot}$), signals can also arise from smaller neutrino bound states that form due to shorter interaction ranges. Since the energy density of these bound states is similar to that of diffuse dark matter halos, they may be probed in dark matter substructure searches \cite{Bonaca:2018fek, VanTilburg:2018ykj,Xiao:2024qay}. Given that the effective mass and number density of neutrinos within these bound states differ from those outside, these bound states may also influence neutrino experiments by producing time-dependent signals as they pass through the Earth (including those aiming to detect the C$\nu$B directly \cite{PTOLEMY:2018jst,PTOLEMY:2019hkd}).
As the parameter space for the long-range forces between neutrinos remains underexplored, these possibilities deserve further investigation.

\acknowledgments


This work was supported by the U.S.~Department of Energy~(DOE), Office of Science, National Quantum Information Science Research Centers, Superconducting Quantum Materials and Systems Center~(SQMS) under Contract No.~DE-AC02-07CH11359. This work is supported in part by the U.S.~National Science Foundation~(NSF) under Grant No.~PHY-2412361.  S.R.~is also supported by the Simons Investigator Grant No.~827042, and by the~DOE under a QuantISED grant for MAGIS. D.E.K.~is also supported by the Simons Investigator Grant No.~144924.  We acknowledge the use of the public cosmological codes \texttt{CAMB}\cite{Lewis:1999bs} and \texttt{CLASS}\cite{Blas:2011rf,Lesgourgues:2011re}.
X.L. thanks Mustafa Amin, Volker Bromm, Anubhav Mathur, Ngan H. Nguyen, Erwin H. Tanin, Huangyu Xiao, Xun-jie Xu and Saiyang Zhang for helpful discussions. X.L. further thanks Anubhav Mathur, Ngan H. Nguyen and Erwin H. Tanin for their comments on the draft.

\appendix

\section{Details of linear perturbation theory}\label{sec:appendix}
In this appendix, we specify the linear perturbation evolution for the C$\nu$B in the presence of the interaction in Eq.~\eqref{lagrangian}. From Eq.~\eqref{lagrangian}, the equations of motion are
\begin{eqnarray}
D_\mu D^\mu\phi + m_\phi^2\phi = -g\bar{\nu}\nu,
\label{EOM_phi}
\\
i\slashed{D}\nu - (m_\nu+g\phi)\nu = 0.
\label{EOM_psi}
\end{eqnarray}
Eq.~\eqref{EOM_phi} means neutrinos will source a vacuum expectation value (vev) of $\phi$, and through Eq.~\eqref{EOM_psi}, this $\phi$ vev changes the mass of neutrino to $ m_\nu +g\phi$. Suppose the distribution function of the C$\nu$B for one neutrino species is given by $F_\nu(p)$. At the homogeneous level
\begin{equation}\label{eq:distribution}
    F_{\nu,0}(p) \approx 2\times \frac{1}{(2\pi)^3} \frac{1}{\exp{(p/T_\nu)}+1},
\end{equation}
here $T_\nu \approx 1.95a^{-1}\text{ K}$. We define the C$\nu$B number density $n_\nu \equiv  \int \mathrm{d}^3p \  F_{\nu}(p)$. The vev of $\bar{\nu}\nu$ is related to the neutrino distribution function by \cite{Esteban_2021,Smirnov:2022sfo,Ghosh:2022nzo,Ghosh:2024qai,Bouley:2022eer} \footnote{Note that the expression Eq.~\eqref{source} works in the regime where $m_\nu+g\phi < 0$. The ``negative'' sign of the effective mass can be absorbed by redefining the fermion state.}
\begin{align}\label{source}
    \braket{\bar{\nu}\nu}(\phi) & = \int \mathrm{d}^3p ~ \frac{m_\nu+ g\phi}{\sqrt{(m_\nu+ g\phi)^2 + p^2}}F_{\nu}(p) \nonumber  \\
    & \approx
    \begin{cases}
 n_\nu   & \ m_\nu+g\phi \gtrsim T_\nu \\[2mm]
 0.46 \frac{m_\nu + g\phi}{T_\nu}  n_\nu& \ m_\nu+g\phi \lesssim T_\nu.
\end{cases}
\end{align}
The energy density of the C$\nu$B can be expressed as 
\begin{equation}
    \rho_\nu(\phi)  = \int \mathrm{d}^3p ~ \sqrt{(m_\nu+ g\phi)^2 + p^2} F_{\nu}(p),
\end{equation}
and the energy density perturbation of the C$\nu$B is defined as $\delta \rho_\nu(\phi,x^\mu)  \equiv \rho_\nu(\phi,x^\mu) - \rho_{\nu,0}$.

We also define the background induced mass of $\phi$ \cite{Esteban_2021}
\begin{align}\label{eq:mphieff}
    m_{\phi,\text{th}}^2 & \equiv  g\frac{\partial \braket{\bar{\nu}\nu}_0}{\partial \phi}(\phi_0) \nonumber \\
    & = g^2 \int \mathrm{d}^3p ~ \frac{p^2}{((m_\nu+ g\phi_0)^2 + p^2)^{3/2}}F_{\nu,0}(p) 
\end{align}
which is non-negligible when the C$\nu$B is relativistic, but becomes very suppressed within less than one Hubble time after the C$\nu$B become non-relativistic. Therefore we will ignore it when analyzing the perturbation evolution of the non-relativistic C$\nu$B.

\subsection{Background evolution}

We separate the homogeneous and perturbed parts of the fields $\phi$ and $F_\nu$ as follows:
\begin{align}\label{perturbation_function1}
    \phi & = \phi_0(\tau) + \delta\phi(x^\mu),
\\\label{perturbation_function2}
F_\nu & = F_{\nu,0}(\tau,p)(1+\Theta(x^\mu,p_\mu)).
\end{align}
Substituting Eqs.~\eqref{perturbation_function1} and \eqref{perturbation_function2} into Eq.~\eqref{EOM_phi} and retaining only the background part yields the equation of motion for the scalar field:
\begin{equation}\label{eq:class1}
       \Ddot{\phi}_0 + 3H\Dot{\phi}_0+ m_\phi^2\phi_0 = -g\braket{\bar{\nu}\nu}_0(\phi_0).
\end{equation}

At the background level, the comoving momentum $q \equiv ap$  is invariant \cite{Anderson:1997un}, so $F_{\nu,0}(q)$ does not evolve over time. Therefore, Eq.~\eqref{eq:distribution} holds even in the presence of $\phi_0$. This behavior can also be derived from the Boltzmann equation \cite{Esteban_2021}. But the neutrino mass now varies with $\phi_0$.

\subsection{Perturbation evolution}
The full equation of motion governing the evolution of $\Theta$ and $\delta \phi$ can be found in Refs.~\cite{Brookfield:2005bz, Bjaelde:2007ki,Esteban_2021}. Following the conventions in Ref.~\cite{Ma1995:astro-ph/9506072v1}, in the synchronous gauge, the Boltzmann equation for the C$\nu$B becomes:
\begin{align}
    & \Theta_0' = -\frac{qk}{\epsilon}\Theta_{1}+\frac{h'}{6}\frac{d\log{F_{\nu,0}}}{d\log{q}},\label{eq:class2}
\\
& \Theta_1' = \frac{qk}{3\epsilon}(\Theta_{0}-2\Theta_{2}) - g\delta\phi  \frac{a^2 (m_\nu +g\phi_0) k}{3q\epsilon}\frac{d\log{F_{\nu,0}}}{d\log{q}},\label{eq:class3}
\\
& \Theta_2' = \frac{qk}{5\epsilon}(2\Theta_{1}-3\Theta_{3})-\frac{h'+6\eta'}{15}\frac{d\log{F_{\nu,0}}}{d\log{q}},\label{eq:class4}
\\
& \Theta_l' = \frac{qk}{\epsilon}(\frac{l}{2l+1}\Theta_{l-1}-\frac{l+1}{2l+1}\Theta_{l+1}),\ \textrm{for}\ l\geq 3.\label{eq:class5}
\end{align}
where $\epsilon = \sqrt{a^2( m_\nu +g\phi_0)^2+q^2}$. Compared to the original Boltzmann equation for the C$\nu$B, an additional term proportional to $g\delta \phi$ appears in Eq.~\eqref{eq:class3} and the neutrino mass depends on $\phi_0$.

The equation of motion for $\delta \phi$ can be derived by substituting Eqs.~\eqref{perturbation_function1} and \eqref{perturbation_function2} into Eq.~\eqref{EOM_phi}:
\begin{align}\label{eq:deltaphi}
    \delta\phi'' + 2aH\delta\phi'+\frac{1}{2}h'\phi'+\left[k^2+a^2(m_\phi^2+m_{\phi,\text{th}}^2)\right]\delta\phi \nonumber \\ = -g\int \mathrm{d}^3 q ~ \frac{ m_\nu +g\phi_0}{\epsilon}F_{\nu,0}(q)\Theta_0(q,k,\tau).
\end{align}

\subsection{Fluid approximation}

Although we have applied Eqs.~\eqref{eq:class1}-\eqref{eq:class5} to the \texttt{CLASS} code by modifying the Boltzmann equations for the ``non-cold relics'', it is difficult to gain physical intuition from these complicated equations of motion. Here we are primarily interested in understanding how long-range forces between neutrinos lead to nonlinear structure formation after the C$\nu$B becomes non-relativistic. This process can be effectively described by treating the C$\nu$B as a non-relativistic fluid. The corresponding simplifications and derivations are discussed below.

First, $\delta \phi$ induced enhancement of perturbations occurs only when the C$\nu$B becomes non-relativistic. Under the assumption of the non-relativistic limit, many background-induced effects become negligible, including $\phi_0$ and $m_{\phi,\text{th}}$. Additionally, the ratio of pressure to energy density of the C$\nu$B becomes much smaller than one.

Second, we will ignore the shear stress and higher multipole moments of the C$\nu$B ($\Theta_l = 0$, for $l\geq 2$). This simplification is sufficient to approximate the suppression of perturbation growth due to the large velocities of the C$\nu$B, which are captured by the first two multipole moments of $\Theta$ \cite{Shoji:2010hm, Green:2021gdc}.

Third, for wavelengths with significant perturbation enhancement, the force mediated by $\phi$ is much stronger than the gravitational interaction on the C$\nu$B. Therefore, we can ignore the scalar perturbations $h$ and $\eta$ when deriving the equations of motion for the C$\nu$B perturbations. In addition, the wavelengths of interest are subhorizon and evolve during the matter-dominating universe.

After applying the aforementioned simplifications, the full Boltzmann equations for the C$\nu$B can be reduced to:
\begin{align}
    \Theta_0' = -\frac{qk}{am_\nu}\Theta_{1},\ 
 \Theta_1' = \frac{qk}{3a m_\nu}\Theta_{0} - g\delta\phi \frac{ak}{3q} \frac{d\log{F_{\nu,0}}}{d\log{q}},\label{eq:class6}
\end{align}
which essentially correspond to the continuity equation and the Euler equation. Following Ref.~\cite{Ma1995:astro-ph/9506072v1}, one can express the fluid properties, such as density fluctuations $\delta_\nu$, divergence of fluid velocity $\theta_\nu$ and sound speed $c_s$, in terms of $\Theta_0$ and $\Theta_1$. In the non-relativistic limit, these are given by:
\begin{align}\label{eq:euler2}
    \delta_\nu \approx \frac{\int_0^\infty \mathrm{d}^3q~ am_\nu  F_{\nu,0}(q)\Theta_0}{\int_0^\infty \mathrm{d}^3q~ am_\nu F_{\nu,0}(q)}, \  \theta_\nu \approx k  
    \frac{\int_0^\infty \mathrm{d}^3q~ q F_{\nu,0}(q)\Theta_1}{\int_0^\infty \mathrm{d}^3q~ am_\nu F_{\nu,0}(q)},
\end{align}
and
\begin{align}
    c_s^2 \approx \frac{\int_0^\infty \mathrm{d}^3q~ \frac{q^2}{3am_\nu}  F_{\nu,0}(q)\Theta_0}{\int_0^\infty \mathrm{d}^3q~ am_\nu  F_{\nu,0}(q)\Theta_0}. \label{eq:class7}
\end{align}
By combining Eqs.~\eqref{eq:class6}-\eqref{eq:class7}, one derives
\begin{equation}\label{neutrino_euler2}
\delta_\nu' = -\theta_\nu, \quad \theta_\nu' = -aH\theta_\nu + c_s^2 k^2\delta_\nu   +k^2\frac{g\delta\phi}{m_\nu}.
\end{equation}
Similarly, Eq.~\eqref{eq:deltaphi} simplifies to
\begin{align}\label{poission2}
     (k/a)^2\delta \phi + m_\phi^2\delta \phi = -g n_{\nu,0} \delta_\nu.
\end{align}
Combining Eqs.~\eqref{neutrino_euler2} and \eqref{poission2} yields the equation of motion for $\delta_\nu$
\begin{equation}\label{delta_eom3}
 \Ddot{\delta}_\nu + 2H\Dot{\delta}_\nu = \frac{3}{2}H^2\left[ \frac{\eta^2k^2}{k^2 +k_\phi^2} f_\nu\delta_\nu - \frac{k^2}{k_{\rm{fs}}^2}\delta_\nu \right].
\end{equation}
This result differs from Eq.~\eqref{neutrino_euler} since we have ignored the gravitational potential during the derivation. In the synchronous gauge, the gravitational potential does not appear explicitly. However, in the conformal Newtonian gauge, one can identify the gravitational potential in the equation of motion for $\Theta_1$, which eventually leads to the extra terms in Eq.~\eqref{neutrino_euler}. Alternatively, instead of using the Boltzmann equation, Eq.~\eqref{neutrino_euler} can also be derived from the energy-momentum conservation equation \cite{Brookfield:2005bz, Bjaelde:2007ki}, upon applying the aforementioned simplifications.

\small
\bibliographystyle{apsrev4-1}
\bibliography{ref.bib}
\end{document}